\documentstyle[aps,graphicx,multicol]{revtex}

\begin{document}

\title{Dislocations in quasicrystals and their interaction with cluster-like
  obstacles}
\draft 
\author{R.\,Mikulla$^{1}$, P.\,Gumbsch$^{2}$, and H.-R.\,Trebin$^{1}$}
\address{$^{1}$Institut f\"ur Theoretische und Angewandte Physik der
  Universit\"at Stuttgart, D-70550 Stuttgart, Germany}
\address{$^{2}$Max-Planck-Institut f\"ur Metallforschung,
  Seestra{\ss}e 92, D-70174 Stuttgart, Germany}

\date{\today}
\maketitle
\begin{abstract}  
  A dislocation moving through a quasicrystal leaves in its wake a
  fault denoted phason wall. For a two-dimensional model quasicrystal
  the disregistry energy of this phason wall is studied to determine
  possible Burgers vectors of the quasicrystalline structure. Unlike
  periodic crystals, the disregistry energy is an average quantity
  with large fluctuations on the atomic scale.  Therefore the
  dislocation core structure and mobility cannot be linked to this
  quantity e.g.\ by a Peierls-Nabarro model.  Atomistic simulations
  show that dislocation motion is controlled by local obstacles
  inherent to the atomic structure of the quasicrystal.

\end{abstract}
\begin{multicols}{2}

\narrowtext

In the last five years it has become possible to grow
thermodynamically stable quasicrystals of high structural quality and
sizes \cite{tsai96} which have allowed one to measure properties like
plasticity and fracture. The data have initiated a wealth of studies,
in which the influence of quasiperiodicity on physical properties is
being explored. Simultaneously, potential applications of
quasicrystals have been discussed, for example as friction reducing,
oxidation resistant, non-adhesive coatings, as solar converters or for
hydrogen storage~\cite{ames96}.

The structure of quasicrystals is also responsible for special
mechanical properties. On this subject an impressive set of
experimental results is available: i) At room temperature
quasicrystals are hard as silicon and extremely
brittle~\cite{takeuchi91}.  ii) At about $80\%$ of the melting
temperature they become plastically deformable up to $30\%$ and soften
upon straining~\cite{takeuchi93}. iii) Stress relaxation experiments
reveal unusually high values for the activation enthalpy and for the
activation volume~\cite{feuerbacher95}.  iv) For AlMnPd it has been
proven that the deformation is governed by dislocation
motion~\cite{wollgarten93}.

In this letter we provide atomistic insight into the mechanical
behavior of quasicrystals. It will turn out that two structural
features govern the motion of dislocations: the faults in the wake of
dislocations and cluster-like obstacles in the glide planes.


Quasiperiodic structures can be obtained as cuts of an irrationally
oriented $n$-dimensional hyperplane (physical space) through a
$d$-dimensional periodic crystal~\cite{bak86}, where $d > n$ is the
number of incommensurate length scales~\cite{acceptance}. Due to the
periodicity of the hyper-crystal, dislocations also exist in
quasicrystals~\cite{bohsung89}. The $d$-dimensional Burgers vectors
carry an $n$-dimensional component $b^{||}$ in physical space and a
$(d-n)$-dimensional component $b^{\perp}$ in the complementing
orthogonal space. A moving dislocation in a two-dimensional
quasicrystal leaves in its wake a phason-wall, separating two areas
with a phase difference $b^{\perp}$, which gives rise to faults in the
pattern and therefore differences in the atomic environments compared
to the perfect quasicrystal. Atomic jumps move these faults and the
wall may consequently vanish by diffusive processes.

\begin{figure}[htb]
  \begin{center} 
    \leavevmode 
    \includegraphics[width=8.4cm]{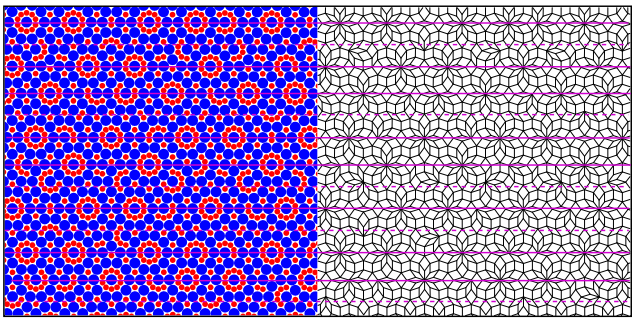}
    \vspace{5pt}
    \caption{Binary tiling obtained from the T\"ubingen triangle tiling
      by decoration. Left: Atoms are displayed as disks (atom
      representation). Right: Bonds are drawn between different atomic
      species (bond representation). The ten-pronged stars represent
      tightly bound clusters and are arranged on families of planes
      (full lines). The easy planes (broken line) run in between the
      full lines of large separation.}
    \label{btlin}
  \end{center} 
\end{figure}

Recently, we performed numerical simulations of plastic deformation
\cite{mikulla95,mikulla95a} on a two-dimensional binary model
quasicrystal displayed in Fig.~\ref{btlin}. We observed that the
phason-wall remained sharp at low temperatures. But close to the
melting temperature it was broadened by strain-induced diffusion and
developed into a several bond-lengths wide glide zone, on which
secondary dislocation dipoles were forming. Since this feature may be
the central mechanism for the strain softening, we further study the
phason-wall here.  We calculate and discuss the ``disregistry
energy'', which is the potential energy necessary to shift one half of
a quasicrystal rigidly over the other along a glide plane (line in two
dimensions). Based on the smallest displacement that provides a
minimal disregistry energy, we construct a generic dislocation by
applying a kink--like displacement field. We then investigate the
dislocation core structure, the friction force impeding dislocation
motion, and the (Peierls) stress to overcome it.

We use the same binary model quasicrystal as in previous shear
simulations~\cite{mikulla95,mikulla95a}. The model originates from the
decagonal T\"ubingen tiling \cite{baake90} of the plane by acute and
flat triangles. The vertices are occupied by large (A) atoms. Small
(B) atoms are placed in the interior of the acute triangles. In
Fig.~\ref{btlin} (left) the system is depicted in the atom
representation, in Fig.~\ref{btlin} (right) in the bond
representation, where next neighbors of different atomic type are
connected. The model is stabilised by Lennard-Jones potentials. The
potential depth is taken as 1 (LJ-unit) for AA and BB bonds, and as 2
for AB bonds in order to avoid phase separation. The LJ unit of length
is one AB bond. 

A striking feature of the structure are concentric rings of ten small
and ten large atoms. Due to the large number of energetically
favorable AB-bonds these rings are tightly bound and resemble the
cluster structure of real systems \cite{janot97}. The simple model
used here therefore reproduces the two most important characteristics
of real quasicrystals: namely quasiperiodicity and clusters.  The
clusters are arranged on families of parallel lines of two separations
which form a Fibonacci sequence. The space in between the large
separation is traversed by a straight line (dashed line in
Fig.~\ref{btlin}) which does not touch any clusters. This line is
characterised by a low surface energy \cite{mikulla96} and serves as
an ``easy'' path for dislocations \cite{mikulla95}. Equivalent
families of lines result from rotations by 36$^{\circ}$.

\begin{figure}[htb]
  \begin{center} 
    \leavevmode
    \includegraphics[width=8.4cm]{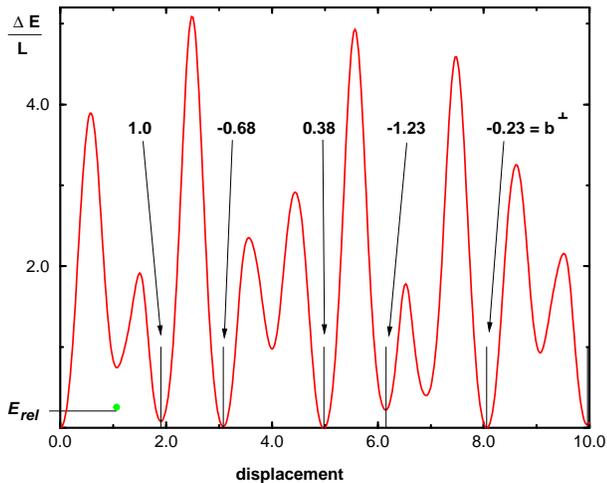}
    \caption{Disregistry energy along the easy plane of
    Fig.~\ref{btlin}. For each minimum the orthogonal component
    $b^{\perp}$ of the Burgers vector is indicated. $E_{rel}$ is the
    energy for the relaxed phason wall corresponding to the second
    minimum.
Lengths and energies are in LJ units.}
    \label{quasidis}
  \end{center} 
\end{figure}

The unrelaxed disregistry energy along such a plane is displayed in
Fig.~\ref{quasidis} (in LJ units). For periodic crystals this energy
is a periodic, approximately sinusoidal curve with zero fault energy
at multiples of the lattice vectors. For our model quasicrystal the
disregistry energy is quasiperiodic, because the spectral
decomposition for the spatial frequencies shows the values 1 and the
inverse golden mean $\tau^{-1} = (\sqrt{5}-1)/2$.

There are two classes of local minima in the disregistry energy.
Those of the first class are very deep, with values of the order
of $10^{-4}$ (vertical thin lines in Fig.~\ref{quasidis}).  The
positions of these minima are projections $b^{||}$ of low indexed
hyper-crystal lattice vectors $b$, whose orthogonal component is
within the acceptance domain of the triangle tiling.  Therefore they
are tiling vectors and represent the primary Burgers vectors of our
model. The small but finite value of the minimum is the energy of the
phason wall for phase difference $b^{\perp}$.

In between there are minima of a second class. Their corresponding
values $b^{\perp}$ are outside the acceptance domain, and the lengths
$b^{||}$ do not belong to the triangle tiling. The energies of the
minima are monotonically (but not linearly) correlated with
$|b^{\perp}|$. The disregistry energies of the second class of minima
are therefore much higher, of the order of 1. Elastic relaxation
reduces this value to $E_{ph} \approx \frac{1}{3}$.

In principle, the $x$-axis is densely covered with projections
$b^{||}$ of lattice vectors $b$. Therefore, a dislocation always can
split into others of smaller Burgers vectors $b^{||}$. However, these
necessarily have larger $b^{\perp}$ components and consequently lead
to a high disregistry energy. Such dislocations of high phasonic component will 
therefore be connected by a high energy phason wall which will
prevent them from splitting. 

Since a moving dislocation necessarily creates a phason wall with
finite phase difference $b^{\perp}$ and finite disregistry energy, it
is interesting to note that such a fault will be a preferred region
for the generation of further dislocations, which then reduce the
disregistry energy.

In periodic crystals a Frank-Read~\cite{frank50} source produces a
sequence of dislocations with the same Burgers vector. In a
quasicrystal such a source would accumulate a phase difference $n
\cdot b^{\perp}$, $n$ being an integer, in the glide plane of the
emitted dislocations. As the stacking fault energy rapidly increases
with $b^{\perp}$, the source cannot operate continuously with a
constant Burgers vector. Instead the dislocation should split into two
with smaller $b^{||}$ and higher $b^{\perp}$. Only $b^{||}$ components
which correspond to the low lying minima in Fig. \ref{quasidis} should
actually be emitted, while the part of the initial Burgers vector
which would create the high energy phason-wall should be left at the
location of the source~\cite{explanat1,explanat2}. Consequently, we
predict that a Frank-Read source in a quasicrystal will not emit
identical Burgers vectors but Burgers vectors of varying length
$b^{||}$. It is obvious that the source will also have to emit Burgers
vectors with larger $b{^\perp}$ components. This result is in
agreement with recent experimental observations \cite{rosenfeld95}
that the probability of dislocations with larger $b^{\perp}$
components increases with plastic strain.

Due to the homogeneity in local surroundings, in periodic crystals the
(global) disregistry energy resembles the local potential for a
dislocation~\cite{vitek75} and hence has been successfully applied by
Peierls~\cite{peierls40} and Nabarro~\cite{nabarro47} for a model of
the dislocation core structure. In quasicrystals there are many
different atomic neighborhoods. The disregistry energy is an average
over the total glide plane. The large local fluctuations can
invalidate any reasoning about the dislocation core structure and
mobility drawn in analogy to periodic crystals.  This observation
is explored by direct atomistic simulations of dislocation cores.
 
In physical space a dislocation with phason wall was constructed by
imposing a kink-like displacement field of total displacement
$b^{||}_{2} = 1.9$ (see Fig.~\ref{quasidis}) on a long slab of the
binary model. The slab had a total length of 300 AB interatomic
spacings and an aspect ratio of 5. The dislocation was placed on a
easy glide plane in the center of the sample.  After relaxation we
imposed homogeneous shear strains between 1$\%$ and 3$\%$. To fix the
deformation on the upper and lower boundary, the atoms were arrested in
a strip four AB interatomic spacings wide. The evolution of the system
then was studied with a constant energy molecular dynamics procedure
from an initial temperature of $10^{-6}$ times the melting
temperature.  We calculated the displacement field with respect
to the defect free initial configuration parallel to the slip plane
and analyzed snapshots of the atomic sites taken from the simulation.

\begin{figure}[htb]
  \begin{center} 
    \leavevmode 
    \includegraphics[width=8.4cm]{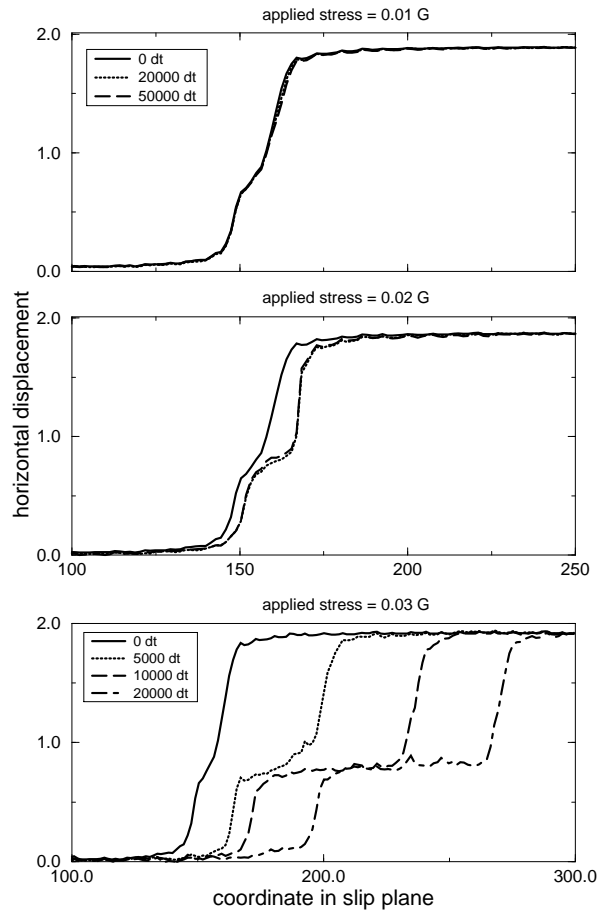}
    \caption{Displacements parallel to the slip plane of a dislocation
      at an applied shear strain of $1\%$, $2\%$ and $3\%$ and for
      different times.}
    \label{disp2}
  \end{center} 
\end{figure}

Fig.~\ref{disp2} shows the horizontal components of the displacement
field parallel to the slip plane for shear deformations of $1\%$,
$2\%$ and $3\%$ at different timesteps. At applied strains of $1\%$
and $2\%$, there is no motion of the dislocation as a whole. Instead,
it remains sessile and splits into two dislocations with
$b^{||}_{1}=1.1$ and $b^{||}_{2}-b^{||}_{1} = 0.8$, which corresponds
to the first secondary minimum in Fig.~\ref{quasidis}. An increase of
the deformation from $2\%$ to $2.5\%$ doubles the separation of the
two dislocations. This observation is a first indication that \textit{not} the
stacking fault energy, i.e.\ the attractive force between the cores of
the partials, but local obstacles dominate their separation.

An increase of the applied shear deformation to a value of $3\%$ leads
to dislocation motion over the total length of the strip (third
diagram in Fig.~\ref{disp2}). After 5000 time steps the dislocation is
split in two dislocations similar to the simulation at $2\%$ shear.
Both then proceed individually with varying velocity and an
oscillating splitting length until the dislocation reaches the
boundary of the system.

Our simulation therefore suggests a value of the Peierls stress
between $2.5\%$ and $3\%$ of the shear modulus G. It is the same
order of magnitude as the value calculated for BCC transition
metals~\cite{vitek75} which show brittle cleavage fracture at low
temperature.  The Peierls stress for a periodic triangular lattice
with a similar interaction model is about $100$ times
lower~\cite{zhou94}. Thus our calculations result in a physically
reasonable value of the Peierls stress for a brittle material like a
quasicrystal.
 
Again we conclude that local obstacles on the glide plane dominate the
dislocation motion since both dislocations moved neither at constant
velocity nor with constant separation. In the following we will
illuminate the microscopic nature of the {\em structure intrinsic}
obstacles observed in our system.

\begin{figure}[htb]
  \begin{center} 
    \leavevmode 
    \vspace{5pt}
    \includegraphics[width=8.4cm]{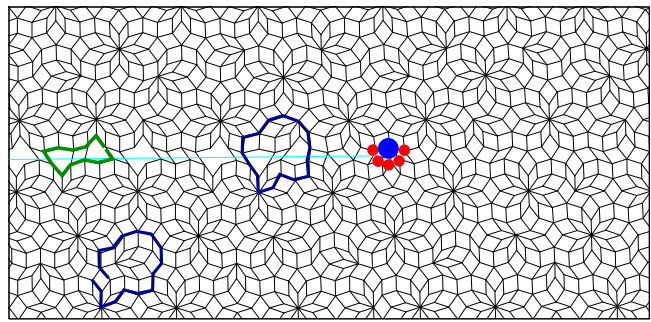}
    \caption{Section of a configuration obtained after 50000 time
      steps at a shear strain of 2$\%$. Both partials are framed by
      Burgers circuits. The Burgers circuit around the leading partial
      is repeated in the undisturbed material below to visualize the
      Burgers vector. The atoms of the environment which acts as an
      obstacle for dislocation motion are drawn as disks. The glide
      plane is marked by a line.}
      \label{obstac} 
    \end{center}
\end{figure}

All simulations performed in this context indicate that the atomic
environment highlighted in Fig.~\ref{obstac} plays the role of a local
obstacle for dislocation motion. Here one large (A) atom on the upper
part is bound to five small (B) atoms below the glide plane.  This is
a highly coordinated atomic cluster, which has the largest
accumulation of bond energy along the "easy" plane. In
Fig.~\ref{obstac} a dislocation stops in front of such an obstacle.
In all calculations up to strains of 2$\%$ an elastic repulsion
between the dislocation core and the atomic environment on the glide
plane with the highest coordination was observed. At higher strains,
in some of our calculations, the leading dislocation core destroys the
obstacle described above and transforms it into a nonplanar
configuration which again is sessile. The dislocation is able to cut
the obstacle only if the strain is increased beyond the Peierls stress.

If the surmounting or cutting of these intrinsic obstacles is the rate
limiting step in dislocation motion, the size of the obstacle may also
explain the high activation volume for dislocation
motion~\cite{feuerbacher95} and has indeed been proposed to do
so~\cite{phDfeuerbacher}. Furthermore, it is worth mentioning that an
atomic cluster which is cut by the first dislocation is a much
weaker obstacle for a second dislocation. Thus the cutting of the
obstacles may even explain the strain softening which has been observed
experimentally~\cite{takeuchi93}.

%
%
Although the binary quasicrystal model used in this paper is very
simplistic, it reflects the most important features of real
quasi\-crystals, i.e.~quasi\-periodicity and clusters and therefore
permits us to reach conclusions relevant and transferable to real
quasicrystals.  The disregistry energy of the phason wall leads to the
identification of true Burgers vectors whose phason wall energies are
at least one order of magnitude lower than the relaxed phason wall
energy of these dislocations of the second kind. These Burgers vectors
have varying length and, consequently a Frank-Read source should
produce more than one dislocation type in a quasicrystal.

Atomistic simulations were performed to calculate the stress which is
necessary to move a dislocation. Its minimal value, the Peierls
stress, is of the order of magnitude of other brittle materials like
BCC transition metals. However, our simulations reveal the dominating
role of highly coordinated atomic environments as {\it structure
  intrinsic} obstacles for the dislocation motion.  The interaction
between dislocations and these obstacles is currently being studied
in more detail with more realistic three-dimensional models.

Financial support from the Deutsche Forschungsgemeinschaft (DFG) under
project Tr 154/6-3 is gratefully acknowledged.

\vspace{-5pt}
\vspace{-20pt}
\vspace{-15pt}
\vspace{-10pt}

\end{multicols}

\begin{references}
\bibitem{explanat1} If a Frank-Read source has a Burgers vector $b =
  (b^{||},b^{\perp}) = (1.9,1)$ which corresponds to the first deep
  minimum in Fig.~\ref{quasidis}, this dislocation can be emitted
  once. However, The second emitted dislocation is then to have a
  Burgers vector $b_e = (1.1,-1.68)$ associated with a low-energy
  phason wall, while the Burgers vector $b_r = (0.8,2.68)$ should
  remain at the location of the source.
\bibitem{explanat2}Although Frank-Read sources have been observed in
  quasicrystals by in-situ transmission
  microscopy~\cite{phDfeuerbacher}, they do not necessarily represent
  the most common sources of dislocations.
\end{references}
\end{document}